\documentclass[10pt, reqno]{amsart}

\usepackage{amssymb} 
\usepackage[english, spanish]{babel}
\usepackage[utf8]{inputenc}
\usepackage{cancel}
\usepackage{multicol}
\usepackage{blindtext}
\usepackage{verbatim}
\usepackage{hyperref}
\usepackage{times}
\usepackage{mathrsfs}

\numberwithin{equation}{section}

\newenvironment{poliabstract}[1]
  {\renewcommand{\abstractname}{#1}\begin{abstract}}
  {\end{abstract}}

\title[Linealización de las Ecuaciones de Campo de Einstein]{Linealización de 
las Ecuaciones de Campo de Einstein}

\author[W. P. \'Alvarez-Samaniego, B. \'Alvarez-Samaniego, D. Moya-Álvarez]{}

\thanks{{\it Fecha}: 27 de septiembre de 2017.}

\email{wpalvarez@uce.edu.ec, alvarezwilson@hotmail.com} 
\email{balvarez@uce.edu.ec, balvarez@impa.br, borys\_yamil@yahoo.com}

\begin{document}
\maketitle 

\centerline{\scshape Wilson P. \'Alvarez-Samaniego}
{\footnotesize
 \centerline{N\'ucleo de Investigadores Cient\'{\i}ficos}
   \centerline{Facultad de Ingenier\'{\i}a, Ciencias F\'{\i}sicas y Matem\'atica}
   \centerline{Universidad Central del Ecuador (UCE)}
   \centerline{Quito, Ecuador}}

\vspace{0.5cm}

\centerline{\scshape Borys \'Alvarez-Samaniego}
{\footnotesize
 \centerline{N\'ucleo de Investigadores Cient\'{\i}ficos}
   \centerline{Facultad de Ingenier\'{\i}a, Ciencias F\'{\i}sicas y Matem\'atica}
   \centerline{Universidad Central del Ecuador (UCE)}
   \centerline{Quito, Ecuador}}

\vspace{0.5cm}

\centerline{\scshape Douglas Moya-\'Alvarez}
{\footnotesize
  \centerline{N\'ucleo de Investigadores Cient\'{\i}ficos}
   \centerline{Facultad de Ingenier\'{\i}a, Ciencias F\'{\i}sicas y Matem\'atica}
   \centerline{Universidad Central del Ecuador (UCE)}
   \centerline{Quito, Ecuador}}


\vspace{1cm}
\selectlanguage{english}

{\centerline{\bf{LINEARIZED EINSTEIN'S FIELD EQUATIONS}}}

\begin{poliabstract}{Abstract}  
From the Einstein field equations, in a weak-field approximation   
and for speeds small compared to the speed of light in vacuum, the 
following system is obtained 
\begin{align*}
   \nabla \times \overrightarrow{E_g} & = 
    -\frac{1}{c} \frac{\partial \overrightarrow{B_g}}{\partial t}, \\
    \nabla \cdot \overrightarrow{E_g} \;\; & \approx -4\pi G\rho_g, \\
   \nabla \times \overrightarrow{B_g} & \approx  
    -\frac{4\pi G}{c^{2}}\overrightarrow{J_g}+ 
	\frac{1}{c}\frac{\partial \overrightarrow{E_g}}{\partial t},  \\
	 \nabla \cdot \overrightarrow{B_g} \;\; & = 0, 
\end{align*}
where $\overrightarrow{E_g}$ is the gravitoelectric field, 
$\overrightarrow{B_g}$ is the gravitomagnetic field, $\overrightarrow{J_g}$ 
is the space-time-mass current density and $\rho_g$ is the space-time-mass density.  
This last gravitoelectromagnetic field system is similar to the Maxwell 
equations, thus showing an analogy between the 
electromagnetic theory and gravitation. 
\end{poliabstract}

\selectlanguage{spanish}

\begin{poliabstract}{Resumen}
A partir de las ecuaciones de campo de Einstein,  en una aproximación de campos débiles 
y para velocidades mucho menores que la velocidad de la luz en el vacío, se 
obtiene el siguiente sistema  
\begin{align*}
   \nabla \times \overrightarrow{E_g} & = 
    -\frac{1}{c} \frac{\partial \overrightarrow{B_g}}{\partial t}, \\
    \nabla \cdot \overrightarrow{E_g} \;\; & \approx -4\pi G\rho_g, \\
   \nabla \times \overrightarrow{B_g} & \approx  
    -\frac{4\pi G}{c^{2}}\overrightarrow{J_g}+ 
	\frac{1}{c}\frac{\partial \overrightarrow{E_g}}{\partial t},  \\
	 \nabla \cdot \overrightarrow{B_g} \;\; & = 0, 
\end{align*}
donde $\overrightarrow{E_g}$ es el campo gravitoeléctrico, 
$\overrightarrow{B_g}$ es el campo gravitomagnético, 
$\overrightarrow{J_g}$ es la densidad de corriente de masa-espacio-tiempo 
y $\rho_g$ es la densidad de masa-espacio-tiempo.  Este conjunto de relaciones 
de campos gravitoelectromagnéticos es análogo a las ecuaciones de Maxwell, 
lo cual muestra una similitud entre la teoría electromagnética y la gravitación.\\ \\
{\textsc{Palabras Claves}}. Campo gravitoeléctrico, campo gravitomagnético, 
ecuaciones de campo de Einstein, ecuaciones de Maxwell, relatividad general.
\end{poliabstract}

\vspace{1cm}

\maketitle
\begin{multicols}{2}
\section{Introducción}\label{Secintr}
De acuerdo a la ley de gravitación universal de Newton y la ley clásica 
de la conservación de la masa, se tiene que
\begin{align}
  &\nabla \cdot \overrightarrow{g}=-4\pi G\rho, \label{eq:Newton}\\
  &\nabla \cdot \overrightarrow{J}+\partial_t\rho =0,  \label{eq:cons}
\end{align}	
donde $\overrightarrow{g}$ es el vector \textit{aceleración de la gravedad}, 
$G$ es la \textit{constante gravitacional de Cavendish}, $\rho $ 
es la \textit{densidad de masa} y $\overrightarrow{J}$ es la 
\textit{densidad de corriente de masa} con 
$\overrightarrow{J}=\rho \overrightarrow{v}$, siendo $\overrightarrow{v}$ la
\textit{velocidad de desplazamiento de la materia}.
Despejando $\rho$ de (\ref{eq:Newton}) se obtiene
\begin{equation*}
  \rho =-\frac{\nabla \cdot \overrightarrow{g}}{4\pi G}.
\end{equation*}	
Reemplazando la expresión anterior en (\ref{eq:cons}), y suponiendo que 
el campo vectorial $\overrightarrow{g}$ es dos veces diferenciable, se ve que
\begin{equation*}
  \nabla \cdot \overrightarrow{J}-\frac{1}{4\pi G}
	 \nabla \cdot \left(\partial_t\overrightarrow{g} \right)=0.
\end{equation*}
Luego,
\begin{equation*}
  \nabla \cdot \left(\overrightarrow{J}-\frac{1}{4\pi G} \partial_t 
  \overrightarrow{g}\right)=0.
\end{equation*}
Usando el Lema de Poincaré, en subconjuntos abiertos contractibles, 	
se tiene que la expresión dentro del paréntesis en la igualdad anterior 
proviene del rotacional de un campo vectorial, que por conveniencia se lo 
escribe como 
\begin{equation*}
  -\frac{c^{2}\overrightarrow{B_g}}{4\pi G}, 
\end{equation*}	
donde $c$ es la \textit{velocidad de la luz en el vacío} y 
$\overrightarrow{B_g}$ se denomina \textit{campo gravitomagnético}.  Luego, 
\begin{equation} \label{eq:rotacional}
  -\nabla \times \Big{(}\frac{c^{2}\overrightarrow{B_g}}{4\pi G}\Big{)}
  =\overrightarrow{J} -\frac{1}{4\pi G} \partial_t \overrightarrow{g}.
\end{equation}
Finalmente,
\begin{equation}\label{eq:Maxwell-Ampere-gravit}
  \nabla \times \overrightarrow{B_g}=-\frac{4\pi G}{c^{2}}\overrightarrow{J}%
  +\frac{1}{c^{2}} \partial_t \overrightarrow{g}.
\end{equation}	
La ecuación (\ref{eq:Maxwell-Ampere-gravit}) es análoga a la Ley de 
Maxwell-Ampère de la Electrodinámica Clásica dada por 
\begin{equation*}\label{eq:Maxwell-Ampere-electr}
  \nabla \times \frac{\overrightarrow{B}}{\mu_0} 
  = \overrightarrow{J_e}+\varepsilon_{0}
  \; \partial_t \overrightarrow{E},
\end{equation*}	
donde $\overrightarrow{B}$ es el \textit{campo de inducción magnético}, 
$\mu_0$ es la \textit{permeabilidad magnética del vacío}, 
$\overrightarrow{J_e}$ es la \textit{densidad de corriente}, 
$\varepsilon_{0}$ es la \textit{permitibidad eléctrica del vacío} y 
$\overrightarrow{E}$ es el \textit{campo eléctrico}.

Se observa ahora  que las ecuaciones gravitacionales (\ref{eq:Newton}) y 
(\ref{eq:Maxwell-Ampere-gravit}), es decir   
\begin{equation}\label{sistema1}
  \left\{
  \begin{array}{ccc}
  \!\!\! \nabla \cdot \overrightarrow{g}  &\!\!=\!\!&  -4\pi G\rho  \\ 
  \!\!\! \nabla \times \overrightarrow{B_g} &\!\!=\!\!&  
         -\frac{4\pi G}{c^{2}}\overrightarrow{J}+
  \frac{1}{c^{2}} \partial_t \overrightarrow{g}
  \end{array}
  \right. \!\!
\end{equation}
son an\'{a}logas a las ecuaciones de Gauss y de Maxwell-Ampère respectivamente 
de la Electrodinámica Clásica
\begin{equation}\label{sistema2}
  \left\{
  \begin{array}{ccc}
  \nabla \cdot \overrightarrow{E} &=& \frac{\rho_e}{\varepsilon _{0}} \\ 
  \nabla \times \overrightarrow{B} &=& \mu_0 \overrightarrow{J_e}+ \frac{1}{c^2} 
  \partial_t \overrightarrow{E},
  \end{array}
  \right.
\end{equation}
donde $\rho_e$ es la \textit{densidad volumétrica de carga} y 
$\mu_0 \varepsilon_{0} = \frac{1}{c^2}$.  

Considerando la existencia de ondas gravitacionales (\cite{Abbott}), se tiene que
\begin{equation}\label{sistema3}
  \left\{
  \begin{array}{ccc}
  \nabla \times \overrightarrow{g} &=& - \partial_t \overrightarrow{B_g}  \\ 
  \nabla \cdot \overrightarrow{B_g} &=& 0.
  \end{array}
  \right.
\end{equation}
Las dos ecuaciones en el sistema (\ref{sistema3}) son similares a la ley de 
Faraday y a la ausencia de monopolos magnéticos de la Electrodinámica Clásica, 
dadas por
\begin{equation}\label{sistema4}
  \left\{
  \begin{array}{ccc}
  \nabla \times \overrightarrow{E} &=&- \partial_t \overrightarrow{B}\\ 
  \nabla \cdot \overrightarrow{B} &=& 0.
  \end{array}
  \right.
\end{equation}	

As\'{\i}, juntando (\ref{sistema1}) y (\ref{sistema3}), los campos 
$\overrightarrow{g}$ y $\overrightarrow{B_g}$ quedan completamente descritos 
en la Mecánica Newtoniana, a nivel no relativista, por el siguiente sistema 
de cuatro ecuaciones tipo Maxwell
\begin{equation} \label{Ec:Max}
   \left\{
	 \begin{array}{ccc}
   \nabla \times \overrightarrow{g} &=& - \partial_t \overrightarrow{B_g} \\
   \nabla \cdot \overrightarrow{g} &=& -4\pi G\rho \\
   \nabla \times \overrightarrow{B_g} &=& -\frac{4\pi G}{c^{2}}\overrightarrow{J}
    +\frac{1}{c^{2}} \partial_t \overrightarrow{g} \\
   \nabla \cdot \overrightarrow{B_g} &=& 0.
	 \end{array}
   \right.
\end{equation}

Aplicando el rotacional a la tercera ecuación de (\ref{Ec:Max}), suponiendo que 
$\overrightarrow{g}$ es dos  veces diferenciable y considerando la 
primera y cuarta ecuaciones de dicho sistema, se obtiene la siguiente 
ecuación hiperbólica para el campo $\overrightarrow{B_g}$:
\begin{equation} \label{Ec:onda1}
  \Delta \overrightarrow{B_g} = \frac{4\pi G}{c^{2}} \nabla \times \overrightarrow{J}
  + \frac{1}{c^2} \partial_t^2 \overrightarrow{B_g}.
\end{equation}
Procediendo de forma similar a lo realizado para obtener (\ref{Ec:onda1}), pero 
aplicando esta vez el rotacional a la primera ecuación de (\ref{Ec:Max}), 
suponiendo ahora que el campo $\overrightarrow{B_g}$ 
es dos  veces diferenciable y tomando en cuenta la segunda y tercera 
ecuaciones de (\ref{Ec:Max}), se consigue nuevamente una ecuación de 
tipo hiperbólico para el campo $\overrightarrow{g}$, dada por
\begin{align} 
  \Delta \overrightarrow{g} = &-4 \pi G \left( \nabla \rho 
  + \frac{1}{c^2} \partial_t \overrightarrow{J} \right) \nonumber \\
  & + \frac{1}{c^2} \partial_t^2 \overrightarrow{g}. \label{Ec:onda2}
\end{align}

La NASA comprob\'{o} experimentalmente el efecto de $\overrightarrow{B_{g}}$ 
(\cite{Ciufolini1, Ciufolini2}). Se colocó un sistema de satélites en órbita 
alrededor de la Tierra, con los ejes de sus respectivos giroscopios apuntando 
hacia una estrella distante de referencia. Debido a que los giroscopios están 
libres de fuerzas externas, sus ejes deberían continuar apuntando hacia la 
estrella por siempre.  Sin embargo, el cambio en la curvatura del espacio, debido 
a la rotación de la Tierra, implica que las direcciones a las que 
apuntan los ejes de los giroscopios deben cambiar con el paso del tiempo. Estos 
cambios de direcciones relativos a la estrella fija de referencia permiten  
medir la variación en la curvatura del espacio-tiempo generado por la rotación 
de la Tierra.

En la Sección \ref{Secaprox} se linealiza las ecuaciones de campo de Einstein 
en la aproximación no relativista para campos gravitatorios débiles, obteniéndose 
así el sistema de relaciones tipo Maxwell dadas en (\ref{ReltipoMax}).  Finalmente, 
en la Sección \ref{Secconcl} se presentan algunas conclusiones del presente 
trabajo.

\section{Aproximación no Relativista para un Campo Gravitacional Débil} \label{Secaprox}
En la aproximación de campos débiles, la métrica del espacio-tiempo, $g_{ik}$, no 
difiere mucho del tensor métrico de la relatividad especial, es decir, de 
la métrica para el espacio pseudo-euclídeo, denotada por $\eta _{ik}$.  Por 
tal motivo, se puede escribir la métrica del espacio-tiempo, en esta aproximación, 
como la suma de la métrica pseudo-euclídea  más una pequeña perturbación 
$h_{ik}$ ($|h_{ik}| \ll |\eta _{ik}|$).  Así, para todo 
$i, k \in \{0, 1, 2, 3\}=: \mathfrak{L}$, se ve que 
\begin{equation} \label{Ec1}
  g_{ik}\approx \eta_{ik}+h_{ik}, 
\end{equation}
donde 
\begin{equation}\label{Ec2}
 \eta_{ik} = 
  \left\{
  \begin{array}{ccl}
  \!\!\! \delta_{ik}  &\!\!\text{, si}\!\!&  i=0 \text{ o } k=0,  \\ 
  \!\!\! -\delta_{ik} &\!\!\text{, si}\!\!&  i,k \neq 0.
  \end{array}
  \right. \!\!
\end{equation}
Considerando los \textit{símbolos de Christoffel} y tomando en cuenta la 
convención de Einstein de suma en los índices repetidos, para 
todo $i, k, l \in \mathfrak{L}$, se tiene que
\begin{equation} \label{Ec3}
  \left\{
  \begin{array}{l}
   \Gamma _{ikl}=\frac{1}{2}\Big(\frac{\partial g_{ik}}{\partial x^{l}}+\frac{
    \partial g_{il}}{\partial x^{k}}-\frac{\partial g_{kl}}{\partial x^{i}}\Big), \\
   \Gamma _{kl}^{i}=g^{im}\Gamma _{mkl}, 
  \end{array}
  \right. \!\! 
\end{equation}  
con  
\begin{equation}  \label{Ec4}
  g^{im}g_{mk} = \delta _{k}^{i}.
\end{equation}
Además, para todo $i, j, k, l \in \mathfrak{L}$, las componentes del 
\textit{tensor de Riemann} (\cite{landau2}) están dadas por
\begin{align*} 
    R^l_{\; ijk}= & \frac{\partial}{\partial x^j} \Gamma^l_{\; ik}
                    -\frac{\partial}{\partial x^k} \Gamma^l_{\; ij}   \\
                  & +\Gamma^l_{\; js} \Gamma^s_{\; ik} 
                    -\Gamma^l_{\; ks} \Gamma^s_{\; ij}.
\end{align*}
Asimismo, para todo $i, j, k, l, m \in \mathfrak{L}$, bajando índices con 
$R_{lijk}=g_{ls} R^s_{\; ijk}$, se ve que
\begin{align} \label{Ec:Rie1}
\!\!\!\!\!\!\!\!\!\!
 R_{iklm}= & \frac{1}{2} \left( \frac{\partial^2 g_{im}}{\partial x^k \partial x^l}
             + \frac{\partial^2 g_{kl}}{\partial x^i \partial x^m}  \right. 
             \nonumber \\
           & \left. - \frac{\partial^2 g_{il}}{\partial x^k \partial x^m}
             - \frac{\partial^2 g_{km}}{\partial x^i \partial x^l} \right) 
             \nonumber \\
           & + g_{np} \big( \Gamma^n_{\; kl} \Gamma^p_{\; im} \nonumber \\
           & - \Gamma^n_{\; km} \Gamma^p_{\; il} \big) .  
\end{align}
Contrayendo dos de los subíndices del tensor dado en (\ref{Ec:Rie1}), se obtiene 
el \textit{tensor de Ricci}, cuyas componentes son
\begin{align} \label{Ec:Rie2} 
  R_{ik} &= g^{lm}R_{limk} \nonumber \\
         &= \frac{\partial \Gamma^l_{\; ik}}{\partial x^l} 
            - \frac{\partial \Gamma^l_{\; il}}{\partial x^k}  \nonumber \\
         & \;\;\;\;  + \Gamma^l_{\; ik} \Gamma^m_{\; lm} 
            - \Gamma^m_{\; il} \Gamma^l_{\; km},
\end{align}
para todo $i, k \in \mathfrak{L}$. De aquí en adelante, se sobrentiende que 
todos los subíndices y superíndices latinos pertenecen al conjunto 
$\mathfrak{L}$ y en cambio los subíndices y superíndices griegos pertenecen 
al conjunto $\{1, 2, 3\}=: \mathfrak{G}$.  Ahora, se mencionan algunas 
propiedades conocidas de los tensores de Riemann y Ricci, que serán usadas 
posteriormente.
\begin{align}
  & R_{ik} = R_{ki},  \label{Ec:R1}\\ 
  & R_{iklm} = -R_{kilm} = -R_{ikml},  \label{Ec:R2}\\
  & R_{iklm} + R_{imkl} + R_{ilmk}=0,  \label{Ec:R3}  \\
  & R^n_{\; ikl;m} + R^n_{\; imk;l} + R^n_{\; ilm;k}=0,  \label{Ec:R4} \\
  & R^l_{\; m;l} = \frac{1}{2}\frac{\partial R}{\partial x^m},  \label{Ec:R5}  \\
  & R = g^{ik} R_{ik},   \label{eqR6}
\end{align}
donde $R$ es el \textit{escalar de curvatura} y el símbolo $;$ en las ecuaciones 
(\ref{Ec:R4}) y (\ref{Ec:R5}) representa la derivada covariante.

Tomando un sistema local de coordenadas tal que la métrica del espacio-tiempo, 
$g_{ik}$, sea diagonal y ya que las perturbaciones, $h_{ik}$, a las componentes 
de la métrica pseudo-euclídea, $\eta_{ik}$, son pequeñas y considerando 
aproximaciones de primer orden en $h_{ik}$, se sigue que 
\begin{equation*} \label{Ec5}
  R_{iklm}\approx \frac{\partial \Gamma _{ikm}}{\partial x^{l}}-
  \frac{\partial \Gamma _{ikl}}{\partial x^{m}}.
\end{equation*}
Así, 
\begin{align} \label{Ec6}
  R_{iklm}\approx  
  & \frac{1}{2}\left[ \frac{\partial ^{2}h_{im}}{\partial x^{k}\partial x^{l}}
    +\frac{\partial ^{2}h_{kl}}{\partial x^{i}\partial x^{m}} \right.
    \nonumber \\
  & \left. -\frac{\partial ^{2}h_{km}}{\partial x^{i}\partial x^{l}}
    -\frac{\partial^{2}h_{il}}{\partial x^{k}\partial x^{m}}\right].
\end{align}
De este modo, las componentes del tensor de Ricci, en esta aproximación 
de primer orden, están dadas por 
\begin{align*} 
  R_{ik} & = g^{lm}R_{limk} \\
         & \approx \eta ^{lm}R_{limk}.
\end{align*}
Luego, 
\begin{align} \label{Ec7}
  R_{ik}  \approx
    & \frac{1}{2}\left[ -\eta^{lm} 
      \frac{\partial^{2} h_{ik}}{\partial x^{l}\partial x^{m}}
      +\frac{\partial^{2} h^l_{\; i}}{\partial x^{k}\partial x^{l}}  
      \nonumber \right.\\
    & \left. +\frac{\partial^{2} h^l_{\; k}}{\partial x^{i} \partial x^{l}}
      -\frac{\partial^{2} h}{\partial x^{i} \partial x^{k}}\right] 
      \nonumber \\
  = & \frac{1}{2}\left[ -\partial ^{l}\partial _{l}h_{ik} 
      +\partial_{k}\partial_{l} h^l_{\; i} \right. 
      \nonumber \\
    & \left. +\partial_{i} \partial_{l} h^l_{\; k}
      -\partial_{i} \partial_{k} h \right],  
\end{align}
donde $h=h^{i}_{\; i}$.

Denotando por $\Phi $ el \textit{potencial gravitatorio}, se tiene 
que la ecuación de Laplace para el espacio libre es
\begin{align} \label{Ec8}
  \Phi _{;i}^{;i} 
  & =  \frac{1}{\sqrt{-g}}
       \frac{\partial}{\partial x^i}
       \left[\sqrt{-g} \; g^{ik}\frac{\partial \Phi}{\partial x^k} \right] 
       \nonumber \\
  & =  0,
\end{align}
donde $g$ es el determinante del tensor métrico del espacio-tiempo.  En el 
caso de campos gravitacionales débiles y usando (\ref{Ec8}), se 
obtiene la siguiente condición de medida para la métrica del 
espacio-tiempo
\begin{equation} \label{Ec9}
 \frac{1}{\sqrt{-g}}\frac{\partial }{\partial x^{i}}
 \left[ \sqrt{-g} \; g^{ik} \right] = 0. 
\end{equation}
Por otro lado, el determinante del tensor métrico, en esta aproximación, 
viene dado por
\begin{align} 
  g & \approx  \; \eta (1 + \eta^{ik} h_{ik}) \nonumber \\
    & \approx  \; \eta (1+h) \nonumber \\
    & = -(1+h),
\end{align}
donde $\eta$ es el determinante del tensor métrico del espacio 
pseudo-euclídeo. Luego, 
\begin{equation} \label{Ec10}
 \sqrt{-g} \approx \sqrt{1+h}
           \approx 1+\frac{h}{2}.
\end{equation}
Así, 
\begin{align*}
  & \frac{\partial }{\partial x^{k}}
    \left[ g^{ik} \left(1+\frac{h}{2} \right)\right] \\
  & = \frac{\partial }{\partial x^{k}} 
    \left[ \left(\eta ^{ik}-\tilde{h}^{ik} \right)
    \left(1+\frac{\tilde{h}}{2} \right) \right] = 0,
\end{align*}
donde $\tilde{h}^{ik} = -h^{ik}$ y $\tilde{h}=\tilde{h}^{i}_{\; i}$.  
Por simplicidad de notación, de aquí en adelante, se escribe $h$ 
para representar $\tilde{h}$.  De la última ecuación y en vista que 
se está considerando aproximaciones de primer orden en $h^{ik}$, 
se obtiene que
\begin{equation*} 
  \frac{\partial }{\partial x^{k}} 
  \left(h^{ik}-\frac{1}{2}\eta ^{ik} h \right) \approx 0.
\end{equation*}
Se define ahora
\begin{equation*} 
  \overline{h}^{ik} := h^{ik}-\frac{1}{2}\eta^{ik} h.
\end{equation*}
Usando las dos últimas expresiones, se sigue que
\begin{equation} \label{Ec11}
  \partial_{k}\overline{h}^{ik} =
  \frac{\partial \overline{h}^{ik}}{\partial x^{k}} \approx 0.
\end{equation}
Ahora, se introducen los siguientes símbolos:
\begin{align}  \label{Ec11a}
  G^{ijk}:= & \frac{1}{2} 
              \left (\overline{h}^{ij,k}-\overline{h}^{ki,j} \right) 
              \nonumber \\
         := & \frac{1}{2} 
              \left( \frac{\partial \overline{h}^{ij}}{\partial x_k}
              - \frac{\partial \overline{h}^{ki}}{\partial x_j}  
              \right) 
              \nonumber \\
          = & \frac{1}{2} 
              \left( \partial^k \overline{h}^{ij}
              - \partial^j \overline{h}^{ki}  \right).
\end{align}
Derivando la última expresión con respecto a la $k$-ésima variable y 
puesto que $\overline{h}$ es dos veces diferenciable, se ve que
\begin{align}  \label{Ec12}
  G^{ijk}_{\;\;\;\; , k} 
  =     & \frac{1}{2}\left( \partial ^{k}\partial _{k} \overline{h}^{ij}
          -\partial_{k}\partial^{j} \overline{h}^{ki} \right)
          \nonumber \\
 \approx & \frac{1}{2} \partial^{k} \partial_{k} \overline{h}^{ij},     
\end{align}
donde en la última expresión se ha usado (\ref{Ec11}) y el hecho que 
se está considerando campos gravitacionales débiles.  Ya que $h$ 
es dos veces diferenciable, se puede escribir 
\begin{equation*}
  \partial_{i} \partial_{k}h
  =\frac{1}{2}\partial_{i}\partial_{k}h
   +\frac{1}{2}\partial_{k}\partial_{i}h.
\end{equation*}
Usando la última ecuación y (\ref{Ec7}), se obtiene 
\begin{align*}
  R_{ik}
  \approx & \frac{1}{2}\left[ -\partial^{l} \partial_{l} h_{ik}
            +\partial_{k} \left(\partial_{l}h^{l}_{\; i}
            -\frac{1}{2}\partial_{i}h \right) \right. \\
          & \left. +\partial_{i}\left( \partial_{l}h^{l}_{\; k}
            -\frac{1}{2}\partial_{k}h \right) \right].
\end{align*}
Los dos términos entre paréntesis en el miembro del lado derecho de la 
última expresión, en esta aproximación de primer orden, son cero pues
\begin{align*}
  \partial_{l} h^{l}_{\; j} - \frac{1}{2} \partial_{j}h
  & =        \partial_{l} \left( g_{mj} h^{lm} \right) 
            -\frac{1}{2} g_{mj} \partial^m h \\
  & \approx  \eta_{mj} \left[ \partial_l h^{lm} 
            -\frac{1}{2} \partial^m h\right] \\
  & \approx  0,      
\end{align*}
donde en la última expresión se ha usado (\ref{Ec11}).  

Procediendo de forma similar, se tiene que las componentes 
contravariantes del tensor de Ricci, en esta aproximación de primer 
orden, están dadas por
\begin{align} \label{Ec13}
  R^{ik}
  & \approx  -\frac{1}{2} \partial^{l} \partial_{l} h^{ik}  
    \nonumber \\
  & =  -\frac{1}{2} \left[ \frac{1}{c^2} 
       \frac{\partial^2}{\partial_t^2} - \Delta \right] h^{ik} 
    \nonumber  \\ 
  & =  \frac{1}{2} \; \Box h^{ik},
\end{align}
donde 
\begin{equation*}
  \Box := \Delta - \frac{1}{c^2} \frac{\partial^2}{\partial_t^2}
\end{equation*}
es el \textit{operador de d'Alembert}.

Además, el escalar de curvatura $R = g_{ik} R^{ik}$,  
en esta aproximación de primer orden y usando (\ref{Ec13}), 
está dado por  
\begin{equation}  \label{Ec14}
    R \approx \eta_{ik} \; \Box \frac{h^{ik}}{2} 
      \approx \frac{1}{2} \; \Box  h.
\end{equation}
Por otro lado, la ecuación de campo de Einstein, sin tomar en 
cuenta el término de la constante cosmológica, es 
\begin{equation} \label{Ec15}
    R^{ik} - \frac{1}{2} g^{ik} R
    =\frac{8 \pi G}{c^{4}} T^{ik},
\end{equation}
donde $T^{ik}$ son las componentes contravariantes del \textit{tensor 
energía-momento}. Reemplazando ahora (\ref{Ec13}) y (\ref{Ec14}) 
en (\ref{Ec15}) y tomando en cuenta el hecho que 
$g^{ik} \approx \eta^{ik}$, se obtiene que
\begin{equation*}
    \frac{1}{2} \; \Box \overline{h}^{ik} 
    \approx \frac{8 \pi G}{c^4} T^{ik}.
\end{equation*}
Usando (\ref{Ec12}) en la última expresión, se sigue que
\begin{equation} \label{Ec16}
    - G^{ikj}_{\;\;\;\; , j} \approx \frac{8 \pi G}{c^4} T^{ik}. 
\end{equation}

{\bf{(i)}} A continuación, se definen las siguientes cantidades:
\begin{equation}  \label{Ec16a}
 G^{00i} := \frac{2}{c^2} E_g^i    
\end{equation}
y
\begin{equation}  \label{Ec16b}
    A^i := \frac{c^2}{4} \overline{h}^{0i},
\end{equation}
donde $E_g^i$ son las componentes del \textit{campo gravitoeléctrico} y $A^i$ 
son las componentes de los \textit{potenciales gravitoelectromagnéticos}.  De 
(\ref{Ec16}), se ve que 
\begin{equation} \label{Ec17}
    G^{00i}_{\;\;\;\; , i} \approx 
    - \frac{8 \pi G}{c^4} T^{00}.
\end{equation}
Además, se conoce que las componentes contravariantes del tensor 
energía-momento vienen dadas por (\cite{landau2})
\begin{equation} \label{Ec18}
   T^{ik}= g^{ik} \mathscr{P} c - (\mathscr{P} c+\mathscr{E})u^{i}u^{k},
\end{equation}
donde $\mathscr{P}$ es la \textit{densidad de momento}, $\mathscr{E}$ 
es la \textit{densidad de energía} y $u^j$ son las componentes 
contravariantes del \textit{cuadrivector velocidad}.  Tomando 
$i=k=0$ en (\ref{Ec18}), se consigue 
\begin{equation}  \label{Ec19}
  T^{00}= g^{00} \mathscr{P} c- (\mathscr{P} c+\mathscr{E})u^{0}u^{0}.
\end{equation}
Por otro lado,  el intervalo de tiempo propio está dado por 
\begin{equation*}
    ds^{2}=g_{00} dx^{0} dx^{0} \approx dx^{0} dx^{0}, 
\end{equation*}
donde en la última relación se ha usado la aproximación de orden 
cero para la métrica del espacio-tiempo.  De la última expresión, 
se observa que
\begin{equation} \label{Ec20}
    u^{0}u^{0} = \frac{dx^{0}}{ds}\frac{dx^{0}}{ds} \approx 1.
\end{equation}
Reemplazando (\ref{Ec20}) en (\ref{Ec19}), se obtiene que 
\begin{align} \label{Ec21}
   T^{00} &  \approx \mathscr{P} c - (\mathscr{P} c+\mathscr{E})\nonumber \\
          & = - \mathscr{E} = - \rho c,
\end{align}
donde $\rho $ es la densidad de masa.  Reemplazando ahora (\ref{Ec16a}) y 
(\ref{Ec21}) en (\ref{Ec17}), se tiene que
\begin{equation*}
  \frac{2}{c^{2}}\frac{\partial E_g^i}{\partial x^i}
  \approx \frac{8 \pi G}{c^{3}} \rho.
\end{equation*}
De la última expresión, se consigue
\begin{equation*}
  \frac{\partial E_g^0}{\partial x^0}
  - \nabla \cdot  \overrightarrow{E_g} 
  \approx \frac{4 \pi G}{c}  \rho.
\end{equation*}
Luego, 
\begin{equation} \label{Ec22}
  \nabla \cdot \overrightarrow{E_g} \approx  -4 \pi G \rho_g,
\end{equation}
donde 
\begin{equation} \label{Ec22a}
  \rho_g := \frac{\rho}{c} 
  - \frac{1}{4 \pi c \; G} \frac{\partial E_g^0}{\partial t}   
\end{equation}
se denomina, de aquí en adelante, \textit{densidad de masa-espacio-tiempo}.

{\bf{(ii)}} Usando ahora (\ref{Ec16a}), se obtiene
\begin{equation*} 
  G^{00i}_{\;\;\;\; , k} - G^{00k}_{\;\;\;\; , i}
  = \frac{2}{c^{2}}\left( \frac{\partial E_g^i}{\partial x^k}-
    \frac{\partial E_g^k}{\partial x^i} \right).
\end{equation*}
Considerando solamente las componentes espaciales en la última 
expresión, usando (\ref{Ec11a}) y el hecho que las componentes 
contravariantes $\overline{h}^{ij}$ son dos veces diferenciables, 
se tiene que 
\begin{align*}
  & \frac{2}{c^{2}} \left[ \frac{\partial E_g^{\alpha }}{\partial x^{\beta}}-
    \frac{\partial E_g^{\beta}}{\partial x^{\alpha}} \right]  \\
  & =-\frac{1}{2}
    \left[ \frac{\partial ^{2}\overline{h}^{0\alpha}}{\partial x^{\beta}
    \partial x^{0}}-\frac{\partial ^{2}\overline{h}^{0\beta }}{\partial
    x^{0}\partial x^{\alpha }} \right].
\end{align*}
Usando ahora (\ref{Ec16b}) y la suposición que las componentes 
contravariantes $\overline{h}^{ij}$ son dos veces diferenciables, se consigue
\begin{equation*}
     \frac{\partial E_g^{\alpha }}{\partial x^{\beta}}-
     \frac{\partial E_g^{\beta}}{\partial x^{\alpha}} 
    = -\frac{\partial}{\partial x^0} 
     \left[ \frac{\partial A^\alpha}{\partial x^\beta}  
     - \frac{\partial A^\beta}{\partial x^\alpha} \right].
\end{equation*}
De la última expresión, se concluye que  
\begin{equation}  \label{Ec23}
  \nabla \times \overrightarrow{E_g}
  =-\frac{1}{c} \frac{\partial \overrightarrow{B_g}}{\partial t}, 
\end{equation}
donde
\begin{equation}  \label{Ec24}
  \overrightarrow{B_g} := \nabla \times \overrightarrow{A}
\end{equation}  
es el campo gravitomagnético.

{\bf{(iii)}} De (\ref{Ec24}) y suponiendo que el campo vectorial 
$\overrightarrow{A}$ es dos veces diferenciable, se deduce que 
\begin{equation}  \label{Ec25}
    \nabla \cdot \overrightarrow{B_g} = 0.
\end{equation}

{\bf{(iv)}} Tomando $i=0$ y $k = \alpha$ en (\ref{Ec16}), se obtiene
\begin{equation} \label{Ec26}
     G^{0\alpha j}_{\;\;\;\; , j} 
     \approx - \frac{8 \pi G}{c^4} T^{0\alpha}. 
\end{equation}
Sin pérdida de generalidad se considera $\alpha=2$ en la última 
expresión, para los otros casos $\alpha \in \{1, 3\}$ se 
procede de forma similar.  Desarrollando la suma en $j$ en 
(\ref{Ec26}), se consigue
\begin{align}  \label{Ec27}
  & G^{020}_{\;\;\;\; , 0} - G^{021}_{\;\;\;\; , 1} 
    - G^{022}_{\;\;\;\; , 2} - G^{023}_{\;\;\;\; , 3}
    \nonumber \\
  & \approx - \frac{8 \pi G}{c^4} T^{02}.
\end{align}
De (\ref{Ec11a}), (\ref{Ec16a}) y (\ref{Ec16b}), se tiene que 
\begin{align*}
   G^{020} & = \frac{1}{2} 
               \left( \frac{\partial \overline{h}^{02}}{\partial x_0}
               - \frac{\partial \overline{h}^{00}}{\partial x_2}  
               \right) \\
           & = - \frac{1}{2} 
               \left( \frac{\partial \overline{h}^{00}}{\partial x_2}
               - \frac{\partial \overline{h}^{02}}{\partial x_0}  
               \right) \\
           & = -G^{002} \\
           & = -\frac{2}{c^2} E_g^2, \\
   G^{021} & = \frac{1}{2} 
               \left( \frac{\partial \overline{h}^{02}}{\partial x_1}
               - \frac{\partial \overline{h}^{01}}{\partial x_2}  
               \right) \\
           & = \frac{2}{c^2} 
               \left( \frac{\partial A^2}{\partial x_1}
               - \frac{\partial A^1}{\partial x_2}  
               \right) \\
           & = \frac{2}{c^2} B_g^3,    \\
   G^{022} & = \frac{1}{2} 
               \left( \frac{\partial \overline{h}^{02}}{\partial x_2}
               - \frac{\partial \overline{h}^{02}}{\partial x_2}  
               \right)    \\
           & = 0 
\end{align*}
y
\begin{align*}
   G^{023} & = \frac{1}{2} 
               \left( \frac{\partial \overline{h}^{02}}{\partial x_3}
               - \frac{\partial \overline{h}^{03}}{\partial x_2}  
               \right) \\
           & = \frac{2}{c^2} 
               \left( \frac{\partial A^2}{\partial x_3}
               - \frac{\partial A^3}{\partial x_2}  
               \right) \\
           & = - \frac{2}{c^2} B_g^1.           
\end{align*}
Reemplazando ahora las últimas cuatro expresiones en (\ref{Ec27}), 
se deduce que
\begin{align*}
    & - \frac{2}{c^3} \frac{\partial E_g^2}{\partial t} 
      - \frac{2}{c^2} \frac{\partial B_g^3}{\partial x}
      + \frac{2}{c^2} \frac{\partial B_g^1}{\partial z} \\
    &  \approx - \frac{8 \pi G}{c^4} T^{02}.
\end{align*}
Luego,
\begin{equation*}
     - \frac{1}{c} \frac{\partial E_g^2}{\partial t} 
     - \frac{\partial B_g^3}{\partial x}
     + \frac{\partial B_g^1}{\partial z} 
     \approx - \frac{4 \pi G}{c^2} T^{02}.
\end{equation*}
Así,
\begin{equation*} 
     \left( \nabla \times \overrightarrow{B_g} \right)^2
     \approx - \frac{4 \pi G}{c^2} T^{02} 
     + \frac{1}{c} \frac{\partial E_g^2}{\partial t}.
\end{equation*}
A continuación, se define la 
\textit{densidad de corriente de masa-espacio-tiempo}, dada por  
\begin{equation*}   
    \overrightarrow{J_g} := 
    \left( T^{01}, T^{02}, T^{03} \right).
\end{equation*}
Entonces, 
\begin{equation*} 
    \left( \nabla \times \overrightarrow{B_g} \right)^2
     \approx - \frac{4 \pi G}{c^2} J_g^2 
     + \frac{1}{c} \frac{\partial E_g^2}{\partial t}.
\end{equation*}
Como se mencionó al inicio de este ítem y procediendo de forma similar a 
lo realizado para el caso $\alpha =2$ arriba, se obtienen expresiones análogas 
a la última relación, pero reemplazando el superíndice 2 por 1 y 3, 
respectivamente.  Por lo tanto, 
\begin{equation} \label{Ec28}
    \nabla \times \overrightarrow{B_g}  \approx  
    -\frac{4\pi G}{c^{2}}\overrightarrow{J_g}+ 
	\frac{1}{c}\frac{\partial \overrightarrow{E_g}}{\partial t}.
\end{equation}
Tomando en consideración (\ref{Ec22}), (\ref{Ec23}), 
(\ref{Ec25}) y (\ref{Ec28}), se obtiene el siguiente sistema 
de relaciones gravitoelectromagnéticas de tipo Maxwell en la 
aproximación de campos débiles no relativista
\begin{equation} \label{ReltipoMax}
\left\{ 
\begin{array}{l c l}
   \nabla \times \overrightarrow{E_g} 
   &=& - \frac{1}{c} \frac{\partial \overrightarrow{B_g}}{\partial t}\\ 
   \nabla \cdot \overrightarrow{E_g}
   &\approx& -4\pi G\rho_g \\ 
   \nabla \times \overrightarrow{B_g}
   &\approx& -\frac{4\pi G}{c^{2}}\overrightarrow{J_g}+
     \frac{1}{c}\frac{\partial \overrightarrow{E_g}}{\partial t} \\ 
   \nabla \cdot \overrightarrow{B_g} &=& 0.
\end{array}
\right. 
\end{equation}
Finalmente, se observa que los sistemas (\ref{Ec:Max}) y 
(\ref{ReltipoMax}) están relacionados a través de la aproximación 
$\overrightarrow{g} \approx c \overrightarrow{E_g}$.

\section{Conclusiones} \label{Secconcl}
{\bf{(1)}} El sistema lineal de relaciones para los campos 
gravitoelectromagnéticos (\ref{ReltipoMax}), en el espacio vacío y 
considerando campos gravitacionales débiles, está dado por
\begin{equation} \label{ReltipoMaxlineal}
\left\{ 
\begin{array}{l c l}
   \nabla \times \overrightarrow{E_g} 
   &=& - \frac{1}{c} \frac{\partial \overrightarrow{B_g}}{\partial t}\\ 
   \nabla \cdot \overrightarrow{E_g}
   &\approx& 0 \\ 
   \nabla \times \overrightarrow{B_g}
   &\approx& \frac{1}{c}\frac{\partial \overrightarrow{E_g}}{\partial t} \\ 
   \nabla \cdot \overrightarrow{B_g} &=& 0.
\end{array}
\right. 
\end{equation}
De (\ref{ReltipoMaxlineal}), se obtienen las siguientes dos 
ecuaciones hiperbólicas
\begin{equation*}
  \Delta \overrightarrow{E_g}  \approx 
  \frac{1}{c^2}\frac{\partial^2 \overrightarrow{E_g}}{\partial t^2} 
\end{equation*}
y
\begin{equation*}
   \Delta \overrightarrow{B_g}  \approx 
  \frac{1}{c^2}\frac{\partial^2 \overrightarrow{B_g}}{\partial t^2},
\end{equation*}
las cuales han sido validadas experimentalmente por medio de la 
detección de ondas gravitacionales (\cite{Abbott}).

{\bf{(2)}} El sistema (\ref{ReltipoMax}) es semejante al sistema de 
ecuaciones de la teoría electromagnética de Maxwell.  Así, se muestra 
una analogía entre la teoría de los campos electromagnéticos y la 
de los campos gravitoelectromagnéticos en la aproximación lineal 
de la relatividad general.

{\bf{(3)}} Usando la aproximación de campos gravitacionales débiles 
se ha obtenido el sistema lineal (\ref{ReltipoMax}) a pesar que 
las ecuaciones de campo de Einstein son altamente no lineales.

{\bf{(4)}} El término 
$- \frac{1}{4 \pi c \; G} \frac{\partial E_g^0}{\partial t}$ en la 
expresión (\ref{Ec22a}) de la densidad de 
masa-espacio-tiempo no es considerado en la teoría clásica de Newton, sin 
embargo, contribuye al valor de la densidad de masa efectiva 
y está asociado a la curvatura del espacio-tiempo en la 
aproximación de campos gravitacionales débiles. 

{\bf{(5)}} Por medio de métodos de cuantización para campos clásicos, 
usuales en la electrodinámica cuántica, se podría pensar en la 
posibilidad de generar un modelo de cuantización del campo 
gravitacional.  Esto será llevado a cabo en un trabajo futuro.


\vspace{2cm}
 \end{multicols}
\end{document}